\documentclass[pdflatex,sn-mathphys-num]{sn-jnl}

\usepackage[english]{babel}
\usepackage[T1]{fontenc}
\usepackage[utf8]{inputenc}
\usepackage{amsfonts}
\usepackage{amsmath}
\usepackage[most]{tcolorbox}
\usepackage{bbding}
\usepackage{fontawesome5}
\usepackage{graphicx}
\usepackage{amssymb}
\usepackage{wasysym}
\usepackage{float}
\usepackage{tikz-cd}
\usepackage{tikz}
\usetikzlibrary{shapes.geometric,arrows.meta,positioning}
\usepackage{physics}
\usepackage{palatino}
\usepackage{adjustbox}
\usepackage{multirow}
\usepackage{enumerate}
\usepackage{array}
\usepackage{amsmath}
\usepackage{braket}
\usepackage{mathtools}
\usepackage{comment}
\usepackage{xcolor}
\usepackage{pifont}
\usepackage{oplotsymbl}
\usepackage{caption}
\usepackage{subfig}
\usepackage{booktabs}
\usepackage{multirow}
\usepackage{tabularx}
\usepackage{geometry}
\usepackage{enumitem}  

\newlist{tabenum}{enumerate}{1}
\setlist[tabenum]{label=\alph*), nosep, leftmargin=*, topsep=0pt, partopsep=0pt}

\begin{document}

\title[Article Title]{Teaching quantum mechanics in secondary schools: A qubit-first learning path to Bell's inequalities}

\author[1]{\fnm{Alberto} \sur{Carlini}}\email{carlinia@liceoking.it}
\author[1]{\fnm{Andrea} \sur{Lanza}}\email{lanzaa@liceoking.it}
\author[1]{\fnm{Dalida} \sur{Monti}}\email{montid@liceoking.it}
\author[1]{\fnm{Anna} \sur{Soso}}\email{sosoa@liceoking.it}
\author[2]{\fnm{Elena} \sur{Losero}}\email{e.losero@inrim.it}
\author*[3]{\fnm{Maddalena} \sur{Mari}}\email{mmari@studenti.uninsubria.it}
\author[3]{\fnm{Matteo} \sur{Oñate Orozco}}\email{monateorozco@uninsubria.it}
\author*[4,5]{\fnm{Maria} \sur{Bondani}}\email{maria.bondani@cnr.it}
\affil[1]{
\orgdiv{Liceo Statale Classico, Scientifico e Sportivo ``M. L. King"}, 
 \postcode{16131}, \city{Genova},
 \country{Italy}}
 \affil[2]{
\orgdiv{Istituto Nazionale di Ricerca Metrologica}, \orgname{QN Metrologia quantistica e nanotecnologie},
 \postcode{10135}, \city{Torino},
 \country{Italy}}
 \affil[3]{
\orgdiv{Dipartimento di Scienza e Alta Tecnologia}, 
\orgname{Università degli Studi dell’Insubria}, 
 \postcode{22100}, \city{Como},
\country{Italy}}
\affil[4]{
\orgdiv{Istituto di fotonica e nanotecnologie}, 
\orgname{Consiglio Nazionale delle Ricerche - CNR-IFN}, 
 \postcode{22100}, \city{Como},
\country{Italy}}
\affil[5]{
\orgdiv{Como Lake Institute of Photonics}, 
\orgname{Università degli Studi dell’Insubria}, 
 \postcode{22100}, \city{Como},
\country{Italy}}
\abstract{Teaching quantum mechanics at the secondary-school level remains a major educational challenge. Traditional historical approaches often present the theory as a sequence of disconnected phenomena, requiring extensive prior knowledge of classical physics before students can appreciate the conceptual structure of quantum physics. In this work, we present the design, implementation, and evaluation of a curriculum based on the two-state (qubit-first) approach, in which the fundamental postulates of quantum mechanics are introduced through mathematically accessible two-level systems and realized with light polarization. The curriculum integrates theoretical instruction, laboratory activities, interactive simulations, and inquiry-based learning into a coherent pathway that progresses from quantum states and superposition to entanglement and the experimental violation of Bell's inequalities. A distinctive feature of the project is the experimental realization of the violation of Bell's inequalities obtained in a newly-established quantum optics laboratory in an Italian high school, where students performed, for the first time in Italy, an authentic Bell test using polarization-entangled photons. The same educational framework was adopted for in-service teacher professional development courses. Learning outcomes were evaluated through mixed methods, both quantitative and qualitative. Preliminary analysis indicates significant improvements in conceptual understanding, confidence in using mathematical formalism, and appreciation of the relationship between theory, experimentation, and the foundations of quantum mechanics. The proposed curriculum demonstrates that a coherent progression grounded in the two-state formalism and supported by authentic laboratory experiences enables secondary-school students to engage more meaningfully with both the conceptual foundations of quantum mechanics and their experimental verification.}
\maketitle


\section{Introduction}\label{sec_intro}
The rapid development of quantum technologies has profoundly transformed both contemporary physics and society \cite{flagship}. Quantum computing, quantum communication, quantum sensing and quantum simulation are no longer confined to research laboratories but are emerging as strategic technologies with important scientific, industrial and economic implications \cite{QTEdu}. This so-called second quantum revolution has generated an increasing demand for a population capable of understanding the fundamental principles underlying quantum technologies, making quantum literacy a new educational priority. Consequently, the teaching of quantum mechanics is progressively shifting from being an advanced topic reserved to university students toward becoming an essential component of secondary education \cite{Merzel2024a}.

Despite this growing interest, introducing quantum mechanics at the high-school level remains a major educational challenge. In many countries, quantum physics is taught only during the final stages of secondary education and often receives limited instructional time \cite{Stadermann2019}. The traditional curriculum usually follows the historical development of twentieth-century physics, presenting quantum mechanics as the solution to the "crisis of classical physics" caused by a bunch of critical experimental results such as black-body radiation, photoelectric effect, atomic spectra and wave-particle duality. While this historical perspective has unquestionable cultural value \cite{Giliberti2024}, it requires students to master a broad range of classical physics topics, including thermodynamics, statistical mechanics and electromagnetism, before they can appreciate the conceptual structure of quantum theory. As a consequence, quantum mechanics is frequently perceived as a collection of disconnected and counterintuitive phenomena rather than as a coherent theoretical framework \cite{Satanassi2025}.

During the past two decades, physics education research has proposed several alternative approaches aimed at making quantum mechanics accessible before university \cite{Krijt2017}. Among the most influential ones, we have the two-state or qubit-first approach, in which the formalism of quantum mechanics is introduced through simple two-level systems, such as spin-$1/2$ particles \cite{McIntyre2022} or polarized single photons \cite{Michelini2000}. This approach allows students to develop the basic fundamental concepts of quantum theory, including state vectors, superposition, measurement, unitary evolution and composite systems, using only elementary linear algebra, complex numbers and probability, mathematical tools that are already available to students during the final years of secondary school \cite{Toth2025,Albert2025,Pospiech2021b,Merzel2024b}. Within this conceptual environment, a wide variety of educational initiatives has emerged. Some focus on quantum computing and quantum information, introducing qubits, quantum gates and simple quantum circuits through online simulators and cloud quantum processors \cite{Satanassi2022,Sutrini2023}. Others exploit the polarization of light to provide an experimentally accessible realization of quantum states \cite{Zuccarini2024}, while quantum cryptography has become an effective context for illustrating the physical consequences of superposition and measurement \cite{Pospiech2021b,Weissman2024}. More recently, educational games \cite{Chiarello2016,qplaylearn,Chiofalo2022,Amabile2026}, interactive simulations \cite{qvis} and analog demonstrators \cite{iqws,Derenzi2025,Derenzi2026,Chiofalo2026} have also been developed to facilitate the understanding of abstract quantum concepts and reduce the cognitive load associated with the formalism. Collectively, these approaches have demonstrated that meaningful learning of quantum mechanics is possible well before university, provided that appropriate pedagogical strategies are adopted \cite{Bondani2022,Bitzenbauer2024}.

Nevertheless, foundational issues, including Einstein–Podolsky–Rosen (EPR) paradox \cite{Einstein1935} and Bell's inequalities \cite{Bell1964,Clauser1969}, are generally regarded as advanced subjects and are rarely included in secondary-school curricula, despite their central role in both the conceptual foundations of quantum mechanics and the operation of many quantum technologies.

From an educational perspective, however, these foundational topics offer a unique opportunity. EPR paradox and Bell's inequalities bring together many of the essential concepts that can be introduced within the two-state formalism, while simultaneously opening discussions about the nature of physical reality, locality and measurement. Moreover, the experimental violation of Bell's inequalities represents one of the most influential experiments in modern physics and forms the conceptual basis of several quantum-information protocols. The award of the 2022 Nobel Prize in Physics to Clauser, Aspect and Zeilinger has further increased public interest in these experiments, making them particularly suitable for motivating students and illustrating the interplay between fundamental research and technological innovation \cite{nobel}.

Building on these considerations, our research group has progressively developed an integrated educational program devoted to quantum science and technology for secondary education. Rather than proposing isolated activities, we adopt a coherent learning progression that begins with the axioms of quantum mechanics applied to two-state systems and develops through light polarization, quantum measurement and entanglement, culminating in the experimental verification of the violation of Bell's inequalities. Throughout the program, mathematical formalism, laboratory activities, interactive demonstrations, educational games and computer simulations are integrated to support conceptual understanding while maintaining a strong connection with authentic scientific practice. The same educational framework has been implemented in professional development courses for in-service teachers and in extracurricular activities for high-school students, allowing continuous refinement of both the curriculum and the associated teaching materials.

A distinctive feature of the program is the realization of a genuine quantum optics laboratory inside an Italian high school through a collaboration between academic researchers and school teachers. Within this laboratory, students performed, for the first time in an Italian secondary school, the experimental measurement of the violation of Bell's inequalities using polarization-entangled photon pairs. The activity was subsequently replicated in a university research laboratory and embedded within a broader program of teacher professional development, providing an opportunity to investigate both the feasibility and the educational value of introducing advanced quantum experiments into secondary education.

In this paper, we present the design principles of the curriculum, describe its implementation, discuss the experimental realization of the Bell test, and report some outcomes of the educational evaluation. We argue that combining the two-state formalism with authentic laboratory experiences provides students with a coherent conceptual understanding of quantum mechanics while simultaneously promoting scientific reasoning, epistemological reflection and familiarity with the principles underlying contemporary quantum technologies.


\section{Educational framework and curriculum design}\label{sec_framework}
The educational pathway presented in this work was developed to address one of the central challenges of quantum education: introducing the conceptual structure of quantum mechanics without requiring the extensive background in classical physics that characterizes the traditional historical approach, but relying only on  elementary mathematical tools already familiar to upper-secondary students.

The curriculum introduces quantum mechanics in its axiomatic structure. Physical systems are described by state vectors belonging to a complex vector space, observables are represented by linear operators, measurements are interpreted probabilistically, and the evolution of isolated systems is described by unitary transformations. Restricting the discussion to two-dimensional Hilbert spaces considerably simplifies the mathematical formalism while preserving the conceptual ingredients that characterize quantum theory.

To achieve our goal, we maintain a continuous interplay between mathematical formalism and physical implementation. Throughout the curriculum every abstract concept is immediately associated with a concrete physical realization. Light polarization provides the experimental implementation of a qubit, polarization states correspond to vectors in a two-dimensional complex space, wave plates implement unitary transformations, while polarizers realize projective measurements. The formal description of polarization allows students to establish a direct correspondence between the mathematical language and laboratory operations, reducing the cognitive gap that often separates formal calculations from experimental observations.

\begin{figure}[ht]
\centering
\includegraphics[width=1\textwidth]{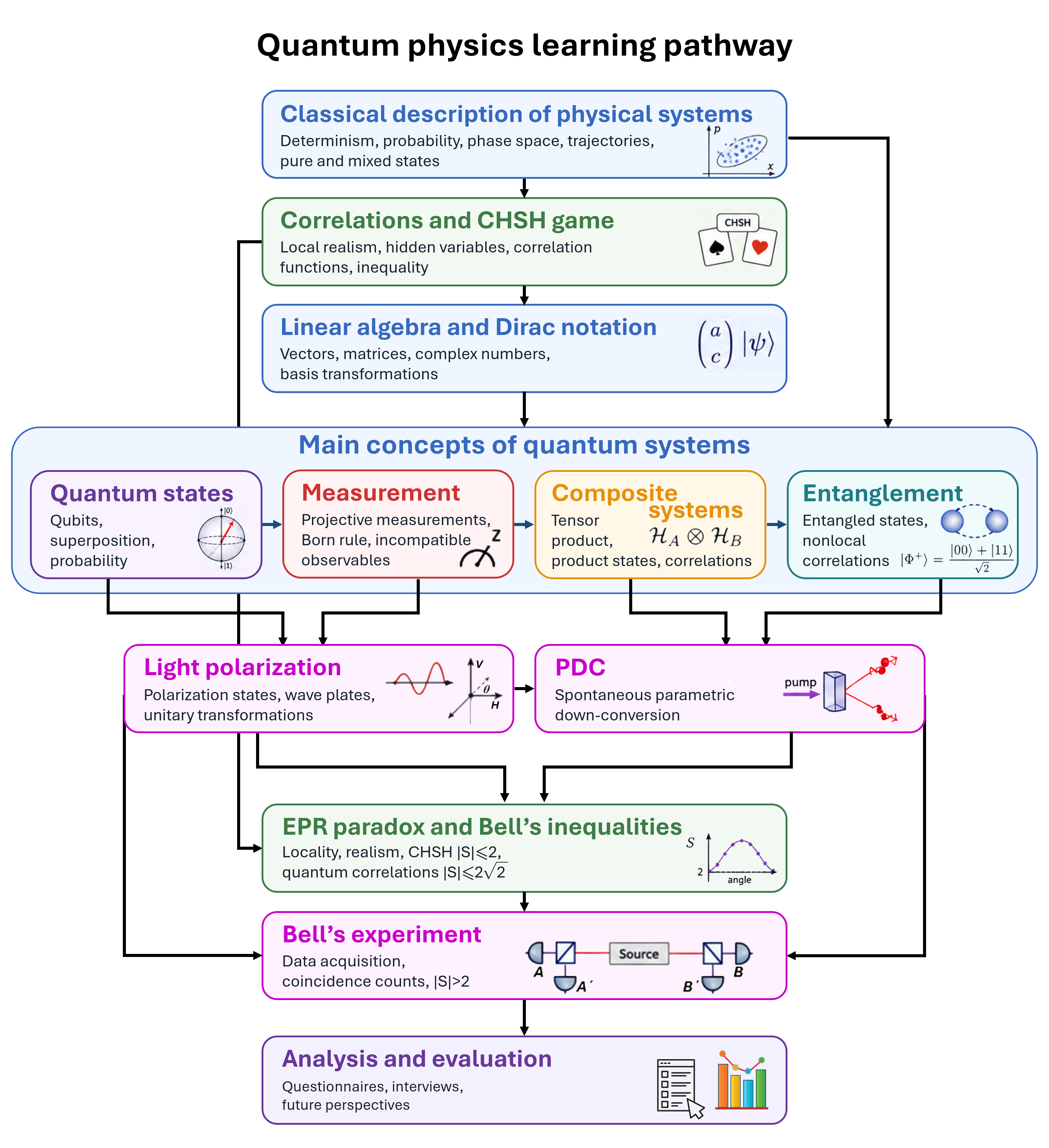}
\caption{Conceptual sequence.}\label{fig_sequence}
\end{figure}

The learning progression follows a carefully designed conceptual sequence (see Fig.~\ref{fig_sequence} and Appendix~\ref{appA} for details). 
The course begins by discussing the classical description of physical systems and the notions of classical state, measurement, probability, and correlations. This initial discussion establishes the conceptual background necessary to appreciate the distinctive features of quantum theory. Basic elements of linear algebra and complex numbers are then introduced as mathematical tools rather than as abstract prerequisites, allowing students to manipulate vectors, matrices, and probability amplitudes within a meaningful physical context.

Once these foundations have been established, the curriculum progressively develops the fundamental postulates of quantum mechanics for two-state systems. Superposition naturally emerges from the vector representation of states, while measurements are introduced as projections onto different bases. The formalism is subsequently extended to composite systems, leading to the concept of entanglement as the impossibility of factorizing the quantum state into that of independent subsystems. This progression enables students to understand entanglement as a direct consequence of the mathematical structure of quantum mechanics rather than as an isolated or mysterious phenomenon \cite{Derenzi2025}. Bell-state generation and Bell-inequality tests emerge naturally from the discussion of composite systems and entanglement.

An important feature of the curriculum is the integration of multiple representational modes. Mathematical derivations are complemented by laboratory demonstrations, computer simulations, graphical representations, and educational games, each contributing different cognitive resources. Rather than replacing mathematical reasoning, these complementary representations provide alternative ways of constructing the same concepts and facilitate transitions between formal and intuitive descriptions. Previous educational experiences within the Italian Quantum Weeks initiative  \cite{iqws,Derenzi2026} and related outreach activities demonstrated that combining different learning environments significantly increases students' engagement while reducing the cognitive load associated with abstract quantum concepts.

Finally, following the practices of contemporary scientific research \cite{Zwickl2014}, students are not expected merely to reproduce predetermined procedures but to participate in authentic experimental activities involving apparatus alignment, optimization of measurement conditions, acquisition of experimental data, statistical analysis, and comparison between theoretical predictions and observations. This inquiry-oriented approach promotes a deeper understanding of the nature of scientific knowledge by highlighting the continuous interaction between theoretical models, experimental evidence, and technological instrumentation.

The same educational framework was adopted for both in-service teacher professional development and extracurricular courses for high-school students. Although the level of discussion and the duration of the activities differed according to the participants, the conceptual progression, the laboratory experiences, and the integration between theory and experimentation remained essentially unchanged. 

The educational program was implemented through parallel courses for in-service secondary-school teachers and upper-secondary students, and was repeated for two subsequent school years in two different Italian locations. In Genoa at the Liceo ``M.L. King”: in-service teachers’ training course (27 hrs, 14 teachers, winter/spring 2025), students' PCTO course (24.5 hrs, 13 students, spring 2025), students' FSL course (30 hrs, 17 students, winter/spring 2025/2026). In Como at the Dipartimento di Scienza e Alta Tecnologia: in-service teachers’ training course (24 hrs, 14 teachers, autumn/winter 2025), students' FSL course (24 hrs, 48 students, winter 2026). 
Although the courses differed in duration and pedagogical emphasis, all followed the same conceptual progression and employed the same educational materials and laboratory activities. It is worth noting that the first courses for teachers and students, held in Genoa in 2025, were taught by academic personnel. Four of the teachers who followed the course became autonomous instructors of the students' FSL course in 2026 in their own school.  

In Appendix~\ref{appA} we present the typical program adopted in the courses for students.


\section{Experimental implementation}

A distinctive feature of the proposed curriculum is the central role assigned to experimental practice. Laboratory activities were not conceived as demonstrations illustrating concepts previously introduced during lectures, but as an essential component of the learning process through which students progressively constructed their understanding of quantum mechanics. 

The experimental program followed the same conceptual progression as the theoretical curriculum. Initial laboratory sessions introduced the experimental tools and measurement procedures that would later be employed in increasingly sophisticated investigations. Subsequent activities gradually transformed these elementary experiments into the complete realization of a Bell test using polarization-entangled photons. In this way, students experienced the experimental verification of Bell's inequalities not as an isolated laboratory exercise but as the culmination of a coherent sequence of scientific investigations.

\subsection{Classical optics experiments}

The first laboratory sessions were devoted to polarization optics using continuous-wave laser radiation and standard optical components to introduce the physical implementation of the mathematical formalism previously developed during the lectures \cite{Pallotta2022}.

Students first characterized the transmission of linear polarizers by quantitatively testing Malus' law for different orientations of the analyzer. Beyond confirming the expected sinusoidal dependence, this activity familiarized participants with optical alignment, intensity measurements, uncertainty estimation and comparison between theoretical predictions and experimental data. The experiment also provided the opportunity to discuss the probabilistic interpretation of polarization measurements, preparing students for the subsequent transition to single-photon experiments.

A second activity investigated the transmission of light through sequences of multiple polarizers, including configurations illustrating the so-called optical analogue of the quantum Zeno effect \cite{zeno,iqws}. By comparing the experimental observations with analytical calculations based on matrix representations of polarization states, students developed confidence in using the same mathematical language to describe both theoretical and experimental situations.

Wave plates were then introduced as optical elements implementing deterministic transformations of polarization states. Their action was analyzed both experimentally and through the two-dimensional vector formalism introduced during the theoretical lectures, emphasizing the correspondence between unitary operators acting on quantum states and optical transformations acting on polarized light. These preliminary experiments established the operational meaning of state preparation, state evolution and projective measurement before the introduction of genuine quantum systems.

\subsection{The Bell-test apparatus}

The following laboratory sessions were dedicated to the detailed analysis of the experimental apparatus designed to demonstrate the violation of Bell's inequalities. Rather than treating the setup as a "black box", students examined the physical function of each individual component and related it to the mathematical formalism previously developed.

The experimental setup was installed inside the optics laboratory of the Liceo Scientifico ``M. L. King'' in Genoa by a collaboration between researchers from CNR-IFN, INRIM and the school physics teachers. The laboratory was funded through the Italian National Recovery and Resilience Plan (PNRR) \cite{PNRR}, making it possible to equip the school with instrumentation comparable to that used in university research laboratories. The same apparatus was subsequently reproduced in the quantum optics laboratories of the University of Insubria.

\begin{figure}[ht]
\centering
\includegraphics[width=1\textwidth]{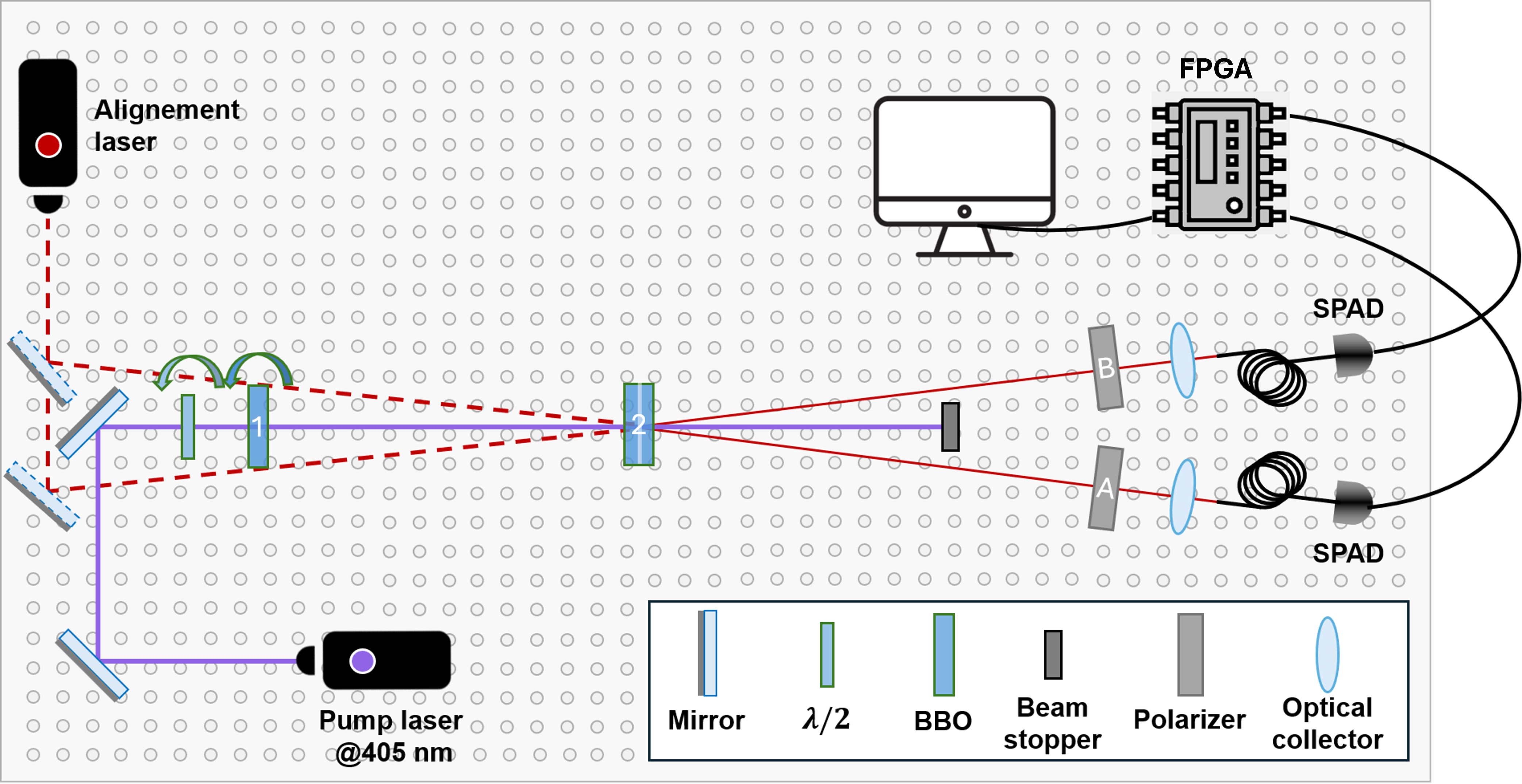}
\vspace{0.2 cm}
\caption{Schematic of the experimental setup. $\lambda /2$: half-wave plate; BBO 1, BBO 2: $\beta$-Barium Borate crystals; SPAD: single photons avalanche detector; FPGA: field programmable gate array.}\label{fig_experiment}
\end{figure}

A schematic of the experimental setup realized in Genoa, as well as some pictures, are reported in Fig. \ref{fig_experiment}. In the experiment design and implementation we started from the work by Beck \cite{Beck2012}, dedicated to junior and senior undergraduates. The goal of Beck's book is to explicate some of the theoretical and experimental aspects of quantum mechanics, proposing several quantum optics experiments dedicated to observe the experimental confirmation of quantum theory, including tests of local realism violating Bell’s inequalities in both Hardy and CHSH formulations.

As remarked during the theoretical lectures, our experiment reproduced, even if in a simplified version, the real experiments performed since 1972, that awarded the Nobel Prize 2022 to their pioneer makers, John Clauser \cite{Clauser1972}, Alain Aspect \cite{Aspect1981} and Anton Zeilinger in \cite{Zeilinger1995}. In particular the specific implementation on the entangled state using a pair of nonlinear crystals reproduces the scheme of Paul Kwiat \cite{Kwiat1999}. 

It is worth noting that any component of the experimental setup admits a precise mathematical representation, which allows for a direct physical implementation of the mathematical concepts introduced through the two-state formalism. We summarize the correspondence in Table~\ref{opt_elem}.

\begin{table}[htbp]
\centering
\caption{Mathematical description of Bell's experiment}
\label{opt_elem}
\begin{tabular}{p{4cm} p{4cm}}
\hline
\textbf{Optical element} & \textbf{Quantum-mechanical role}  \\
\hline
Pump laser & State preparation\\    
BBO 2 crystal pair & Entanglement generation\\
$\lambda/2$ and BBO 1 crystal & State optimization\\   
Polarizers  & Projective measurements\\
SPADs & Single-photon measurements\\
Coincidence electronics  & 	Joint probability estimation\\
\hline
\end{tabular}
\end{table}

The alignement procedure is described in Appendix~\ref{appB}.

\subsection{Experimental realization of Bell's inequalities}

The final laboratory sessions were devoted to the full realization of the Bell test.

The experimental apparatus was set up and aligned by the instructors, while students were involved in the subsequent optimization and measurement procedures. In particular, students followed the optimization protocol by acting on the polarization state of the photons, adjusting the relevant optical elements to maximize the quality of the entangled state, and selecting the measurement bases corresponding to the CHSH formulation of Bell’s inequalities. Coincidence counts for the required combinations of analyzer orientations were collected by the FPGA and elaborated using a LabView program and interface written by Beck \cite{Beck_LV}. The LabView program evaluates the correlation functions and the CHSH parameter, which the students compared with the classical limit.

This activity allowed students to experience the essential steps of an experimental investigation without requiring them to perform the initial alignment and optimization of the light-collection system. They were nevertheless directly engaged in preparing and characterizing the polarization state, acquiring and analyzing the data, and discussing statistical fluctuations and possible sources of systematic uncertainty. The measurements clearly demonstrated a violation of the classical CHSH bound, providing direct experimental evidence of quantum correlations between polarization-entangled photons. Beyond the scientific significance of the result, the activity helped students appreciate how theoretical predictions are connected to experimental procedures, mathematical modelling, data acquisition, and statistical analysis. This connection between theory and experiment constituted one of the principal educational objectives of the curriculum.

\begin{figure}[ht]
\centering
\includegraphics[width=0.95\textwidth]{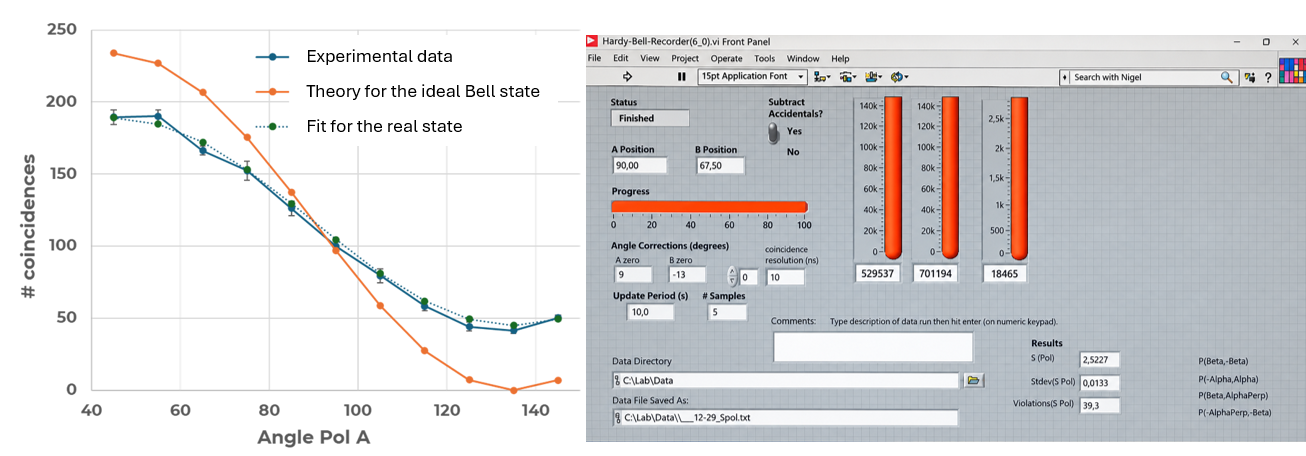}
\vspace{0.2 cm}
\caption{Experimental results obtained with the apparatus in Genoa. Left: preliminary measurements of correlations between the two detectors' outputs as a function of the angle set on one of the polarizers. Right: typical screenshot of the LabView interface. The displyed result is particularly good  $\langle S_{exp} \rangle=2.5227 \pm 0.0133$.}
\label{fig_results}
\end{figure}

The final acquisition, which concluded the educational path, led to a significant violation of the inequalities. In particular, during the first measurement with students we measured: $\langle S_{exp} \rangle=2.32 \pm 0.05$, corresponding to a violation of the classical benchmark $\langle S_{cl} \rangle=2$ by 6.4 standard deviations. Due to the limited amount of time available for the measurements, we had to keep the averaging time short: longer acquisition time would reduce the uncertainty thus increasing the significance of the violation. Moreover, further improving the Bell state optimization led to a value of $\langle S \rangle$ closer to the theoretical value $\langle S_{Bell} \rangle=2 \sqrt{2}$, such as that shown in Fig.~\ref{fig_results}.
Through our experimental setup, we achieved the violation of Bell’s inequalities for the first time in an Italian High School. We obtained similar results in all the acquisitions performed, demonstrating the stability of the system and the repeatability of the experiment, despite the non ideal environmental conditions (e.g. no air conditioning for stable temperature and no isolation frames for the optical table to remove background motion noise).  

Note that in our activity we focused on the realization of Bell’s inequalities violation and we didn’t consider the loophole problems, even if we mentioned them in the theoretical lectures presenting how they were finally solved in 2015 \cite{Hensen2015}, since this aspect is outside the scope of our didactic intervention and requires more advanced experimental tools.


\section{Educational evaluation}\label{sec:eval}

The educational program was evaluated using complementary quantitative and qualitative instruments. In this section, we focus exclusively on the qualitative data collected through semi-structured interviews conducted at the end of the FSL course held in Genoa during winter/spring 2025/2026. The interviews explored students’ perceptions of the course, the concepts they found most challenging or significant, their confidence in using the mathematical formalism, and their reflections on the epistemological implications of quantum mechanics. Particular attention was given to students’ reasoning, learning processes, and the educational value they attributed to engaging with authentic quantum phenomena and experiments.

The overall evaluation strategy also included pre- and post-course questionnaires, discipline-specific conceptual tests, and five-point Likert-scale questionnaires addressing participants’ perceptions of the course, including its difficulty, engagement, confidence in using mathematical tools, appreciation of laboratory activities, and perceived connections between theory and experiment. These instruments were administered but are not analyzed in the present work.  A comprehensive analysis of the results obtained from all the monitoring instruments, including the quantitative questionnaire data, will be presented in a forthcoming paper.

\subsection{Qualitative analysis of students' answers}

The FSL course involved 17 self-selected students and was taught by a group of four teachers (co-authors of the present paper) of the same school as the instructors. As mentioned above, the teachers attended the in-service training course one year earlier, acquiring the skills needed to teach independently. 

To conduct the interviews, students were divided in four focus groups moderated by the instructors. The guiding questions and the rubric used to analyze the answers are described in Appendix~\ref{appC}. We evaluated five dimensions of learning: (D1) the role of mathematics in conceptual understanding, (D2) the perceived cultural value of the EPR paradox and Bell inequalities, (D3) the cognitive impact of confronting non-classical concepts, (D4) disciplinary self-efficacy, and (D5) the overall educational experience. Across the four interview groups, remarkably consistent patterns emerged.

\subsubsection{Mathematics as a tool for conceptual understanding}

The overwhelming majority of students did not perceive the mathematical formalism as an obstacle. Instead, most responses were classified at Level 3 (Support), indicating that mathematics was progressively understood as an interpretative language connecting theoretical models with physical phenomena.

A recurrent theme was that mathematical concepts initially appeared abstract but acquired meaning once they were applied experimentally. As one student explained, {\it "At the beginning I couldn't really understand it, but when we later applied it to physics everything became much clearer."} Similarly, another student observed that {\it "Knowing the mathematics helped me with matrices, polarizers and wave plates. Sometimes, however, it felt like something extra that did not immediately add understanding."}

Several participants argued that additional mathematical exercises or a closer integration between theory and laboratory activities would have facilitated understanding. Others explicitly recognised that the mathematical formalism became meaningful only after the experimental sessions. One participant summarised this process by stating that {\it "The laboratory helped us understand what the professors had been explaining during the theoretical lessons."}

These responses suggest that students progressively shifted from viewing mathematics as a computational requirement toward recognising its explanatory role within quantum theory.

\subsubsection{The cultural relevance of the EPR–Bell problem}

Students generally attributed considerable value to the topics discussed during the course, although the perceived relevance varied. Most responses were classified between Level 2 (Disciplinary interest) and Level 3 (Cultural relevance).

Many students argued that Bell's theorem is particularly important for those interested in physics rather than for the general public. For example, one participant commented, 
{\it "If someone is interested in physics, then yes; otherwise I wouldn't consider it fundamental."} Others adopted a broader perspective, emphasising that the course revealed that physics still contains unresolved conceptual questions. One student remarked, 
{\it "It stimulates you to ask many more questions than if there were already an answer."} 
Another added that quantum mechanics encourages students 
{\it "not to stop at the physics we normally study at school, but to go beyond it."}

Only a minority explicitly argued that Bell's theorem should become part of everyone's scientific culture. Nevertheless, the interviews indicate that students increasingly perceived quantum mechanics as a discipline addressing fundamental questions about scientific knowledge rather than simply introducing new physical phenomena.

\subsubsection{Cognitive impact of Bell's inequality violations}

Perhaps the most significant finding concerns students' reactions to the violation of Bell inequalities. Most participants described an initial sense of surprise, followed by gradual acceptance of a fundamentally different conceptual framework. 
Typical responses correspond to Level 2 (Superficial cognitive conflict) or Level 3 (Conceptual conflict). One student summarised this evolution by saying, 
{\it "At first it felt strange, but then it simply became a new way of seeing things."} 
Another similarly stated, 
{\it "At the beginning I was surprised because it is a completely different way of looking at reality."}

Several students explicitly recognised the conflict between classical intuition and quantum theory without rejecting the latter. One participant observed, 
{\it "It is strange because classical physics is very different. You first have to enter that new way of thinking before you can understand it."} 
Interestingly, no student expressed outright rejection of the experimental conclusions. Instead, conceptual tension appeared to stimulate curiosity. As one interviewee concluded, 
{\it "It is a new world that has to be studied, understood and analyzed."}

These findings suggest that confronting Bell's theorem generated productive cognitive conflict rather than resistance to conceptual change.

\subsubsection{Development of disciplinary self-efficacy}

Nearly all students reported increased confidence in approaching advanced physics, corresponding predominantly to Level 3 (Disciplinary growth).

Rather than claiming complete mastery, students described increased curiosity and willingness to engage with difficult scientific ideas. One participant explained, 
{\it "Having understood topics that are much more difficult than usual made me more motivated."}
Another stated, {\it "Now I feel more immersed in physics. I don't understand everything, but I feel much more involved."}

Students also expressed greater confidence in discussing physics with others. As one remarked,
{\it "I would try to explain to my friends what we have done."}

These responses indicate that the course strengthened students' confidence in learning challenging scientific concepts while simultaneously increasing their motivation to continue studying physics.

\subsubsection{Overall evaluation of the course}

Students consistently evaluated the course positively. Most responses correspond to Level 3 (Positive evaluation), while several reached Level 4 (Transformative experience) because students explicitly described changes in their understanding of physics.

A recurring theme was that the course proved substantially more demanding than initially expected. Nevertheless, this difficulty was interpreted positively. As one student commented,
{\it "At first I thought it would be impossible. Now I would say that it is difficult, but achievable."} Another observed, 
{\it "I expected it to be difficult, and it was, but thanks to this course I discovered concepts that I had never encountered before."}

The experimental activities emerged as one of the strongest positive aspects of the course. Students repeatedly described laboratory work as the moment when the theoretical concepts became meaningful. One participant summarised this relationship by stating, 
{\it "The laboratory was fundamental because you could see everything with your own eyes, use the instruments and take the measurements."}

Several students even suggested extending the course to include additional experimental sessions integrated with the theoretical lessons, reinforcing the central role of experimentation in supporting conceptual understanding.

\subsection{Emerging educational themes}

Across all interviews, four robust themes consistently emerged. First, mathematics was not perceived as a barrier but as a meaningful language once it was continuously connected to experimental evidence. Second, laboratory activities functioned as an essential bridge between abstract formalism and physical interpretation. Third, Bell's theorem generated productive cognitive conflict that promoted conceptual restructuring rather than resistance to quantum ideas. Finally, the course substantially enhanced students' disciplinary self-efficacy by increasing their confidence, curiosity, and willingness to engage with advanced scientific concepts.

Overall, the qualitative evidence suggests that integrating mathematical formalism, authentic laboratory experiences, and explicit discussion of the epistemological implications of quantum mechanics supports not only conceptual learning but also the development of a more sophisticated understanding of scientific knowledge itself. This combination appears to foster both intellectual engagement and confidence in addressing some of the most counterintuitive ideas of modern physics.


\section{Discussion, conclusions and outlook}

The rapid emergence of quantum technologies is creating a growing need for educational approaches capable of introducing quantum mechanics before university. The results of this study suggest that advanced concepts such as quantum measurement, entanglement and Bell's inequalities can be made accessible at the upper-secondary level when they are embedded in a coherent conceptual framework that connects the mathematical formalism, experimental practice and a reflection on the foundations of quantum theory.

The proposed approach contributes to a growing body of research advocating alternatives to traditional historically driven introductions to quantum mechanics. Work by Michelini and collaborators \cite{Michelini2000} has emphasized the value of reconstructing quantum mechanics around its fundamental concepts and of using light polarization as an accessible entry point. Pospiech \cite{Pospiech2021a} has highlighted the role of mathematics in secondary-school quantum education, while Bitzenbauer and collaborators \cite{Bitzenbauer2024} have provided evidence for the educational potential of two-state approaches and Dirac notation. Kohnle's work \cite{Kohnle2015} has demonstrated the value of interactive and visual approaches, whereas Greca and collaborators \cite{Greca2000} have emphasized conceptual change and model-based learning. More recent approaches, including those associated with De Renzi and collaborators \cite{Derenzi2026}, have explored quantum information as a framework for introducing contemporary quantum physics.

The present curriculum shares elements with these approaches but differs in the way they are integrated. The two-state formalism provides a common mathematical language throughout the learning pathway, while polarization optics, hands-on experimentation, educational games and discussion activities provide complementary ways of constructing meaning. Quantum information concepts are consequently not presented primarily as technological applications, but as developments of the underlying structure of quantum mechanics. This distinction is important because it allows technological topics such as quantum communication or computing to be introduced while maintaining the conceptual foundations of the theory as the primary educational objective.

A particularly distinctive aspect is the integration of authentic quantum-optics research within secondary education. The installation of a quantum optics laboratory in the school enabled students to progress from theoretical descriptions of quantum states and measurements to the acquisition and analysis of experimental data from polarization-entangled photons. Bell's test therefore served not simply as a demonstration, but as an authentic scientific investigation in which mathematical predictions, experimental procedures and statistical analysis were brought together. This experience suggests that advanced experiments traditionally associated with university laboratories can become feasible at secondary-school level when they are embedded in an appropriate conceptual and educational progression.

The preliminary evaluation of the activity provides encouraging evidence for this approach, suggesting in particular that introducing the two-state mathematical formalism explicitly does not necessarily increase the difficulty of learning quantum mechanics. When vectors, matrices and the Dirac notation are continuously connected to physical systems and measurements, mathematics can, in fact,  function as an explanatory language through which students interpret experimental phenomena. 

The parallel implementation with teachers is also relevant to the sustainability of the approach. Some of the school teachers, who attended the training course in Genoa in the 2024/2025 school year, held the same FSL course for students in the 2025/2025 school year. This demonstrates that the transfer of expertise from researchers to teachers is feasible and necessary, if advanced quantum education is to move beyond isolated projects and become part of regular school practice.

Several limitations should nevertheless be considered. The present implementations involved a relatively small number of motivated participants and was conducted in an extracurricular setting, where substantially more time and laboratory resources could be devoted to the subject than are normally available in the standard physics curriculum. The present evaluation should therefore be regarded as an initial assessment of the feasibility and educational potential of the approach. A more systematic quantitative analysis of the questionnaire data, based on substantially larger groups of participants and repeated implementations of the programme, will be presented in a forthcoming study. This analysis will allow students' learning outcomes, perceptions and attitudes to be examined with greater statistical robustness and will provide a stronger basis for comparing different implementations and identifying the factors that most strongly contribute to the effectiveness of the proposed approach.

The experimental component can likewise be developed further. Students could be involved more extensively in alignment procedures, preparation and measurement of different polarization states, and the complete analysis of raw coincidence data. These activities would increase their responsibility for the experimental process and strengthen the connection between theoretical predictions and measured results. Technical improvements to the apparatus, including automated polarization rotation and improved preparation of the pump polarization, could further increase measurement efficiency and data quality. The platform could subsequently be extended to more advanced activities, such as quantum state tomography.

The long-term objective is to develop the laboratory and curriculum as an open educational resource for a broader community of schools and teachers. The participation of teachers in the initial implementation provides a basis for organizing future editions independently, while opening the laboratory to students and teachers from other schools could support an open-science model of quantum education. Such a model would allow advanced experimental resources to be shared beyond the school in which they were initially developed and could facilitate the formation of a wider community of practice.

The curriculum can also be extended beyond the topics addressed in the present implementation. Because the two-state formalism provides a common conceptual and mathematical foundation, subsequent modules can introduce quantum communication, quantum computing, quantum sensing and quantum metrology without abandoning the framework already established by students. The increasing availability of affordable quantum-optical instrumentation, educational platforms and cloud-access quantum processors makes such extensions increasingly realistic.

Overall, the main contribution of this work is not a single teaching activity or experimental demonstration, but the integration of these elements into a coherent pathway from fundamental quantum concepts to authentic experimental investigation. The experience shows that meaningful quantum education at secondary-school level can combine mathematical rigor, experimental authenticity and conceptual reflection without reducing quantum mechanics to either a purely qualitative subject or a collection of technological applications. In this sense, the proposed approach represents a possible model for developing quantum literacy that enables students not only to become familiar with emerging quantum technologies, but also to understand the physical principles and scientific reasoning on which they are based.


\section*{Declarations}

\subsection*{Funding}
NQSTI (National Quantum Science and Technology Institute), FUTURA
La Scuola per l’Italia di Domani, the Italian Ministero dell’Universit\`a e della
Ricerca and Ministero dell’Istruzione e del Merito, funded by the PNRR (Piano Nazionale
di Ripresa e Resilienza) and by the European Union’s Next Generation EU; QUID
(QUantum Italy Deployment) and EQUO (European QUantum ecOsystems), funded by
the European Commission in the Digital Europe Programme under the grant agreements
Nos. 101091408 and 101091561.

\subsection*{Ethics approval and consent to participate}
Not applicable.

\subsection*{Conflict of interest/Competing interests}
Not applicable.

\subsection*{Dual publication}
The results/data/figures in this manuscript have not been published elsewhere, nor are they under consideration by another publisher.

\subsection*{Authorship}
I confirm the corresponding author has read the journal policies and submit this manuscript in accordance with those policies.

\subsection*{Third party material}
All the material is owned by the authors and no permissions are required.

\subsection*{Availability of data and materials}
The data that support the findings of this study are available upon request and with the permission of Liceo ``M.L. King'', Genoa, and CNR-IFN, Como.

\subsection*{Code availability}
Not applicable.

\subsection*{Authors' contributions} 
A.C., E.L. and M.B. wrote the main manuscript text; A.C., E.L., M.O.O. and M.B. installed the experiment in Genoa; M.M. and M.B. installed the experiment in Como; A.C., A.L., D.M., A.S. and M.B. delivered the lectures; all authors participated in the experimental activities; A.C., A.L., D.M., A.S., M.M. prepared, administered and analyzed the questionnaires; all authors reviewed the manuscript.

\subsection*{Acknowledgements}

A.C. and M.B. are grateful to M. Beck for his invaluable support. A.C., M.M. and M.B. thank M. Malgieri for discussions and help in elaborating the questionnaires.

\begin{appendices}

\section{Educational pathway}\label{appA}

\begin{figure}[ht]
\includegraphics[width=1\textwidth]{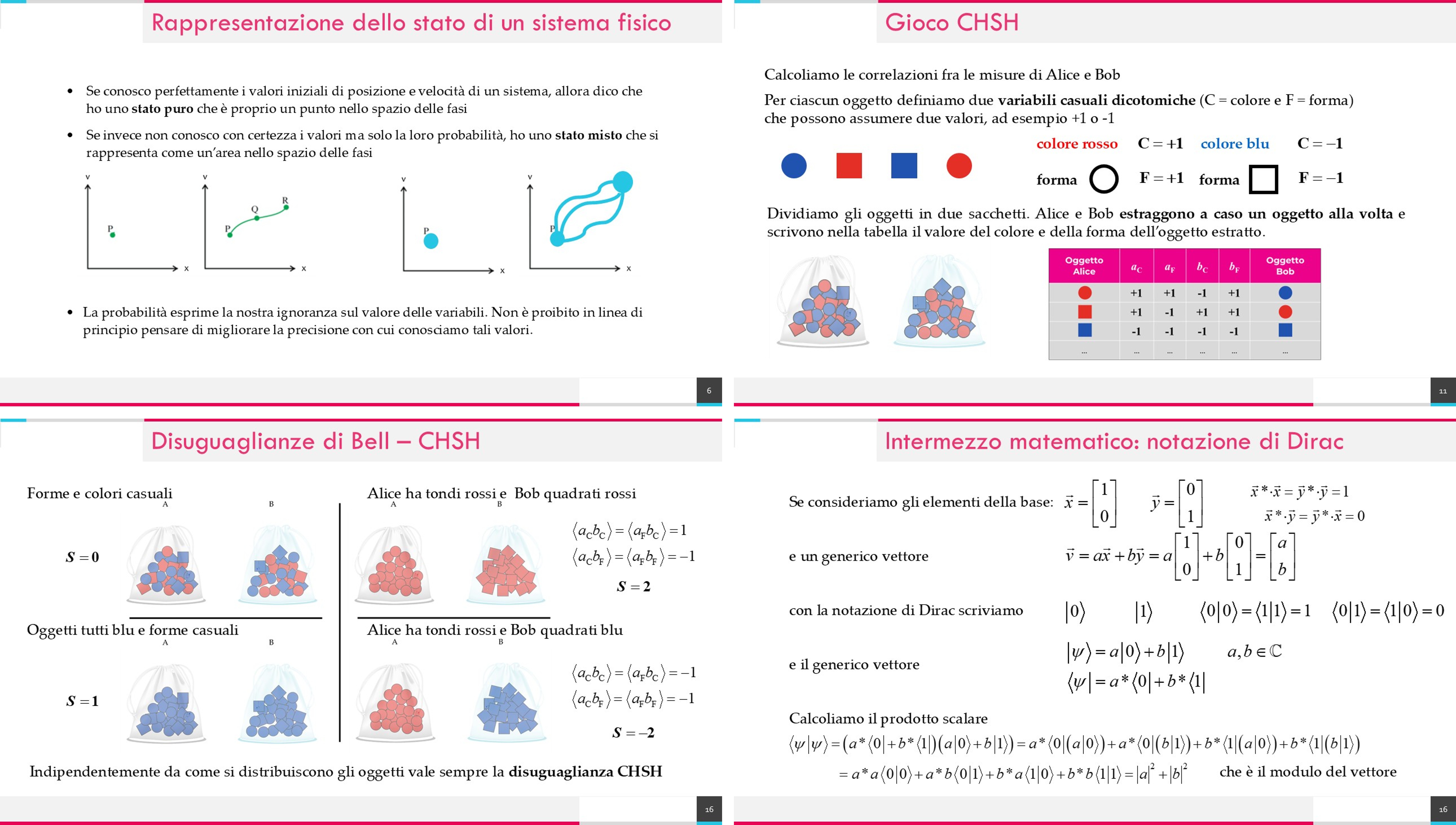}
\vspace{0.2 cm}
\caption{Example of slides extracted from the first two lectures.}\label{L1_2}
\end{figure}

\begin{figure}[ht]
\includegraphics[width=1\textwidth]{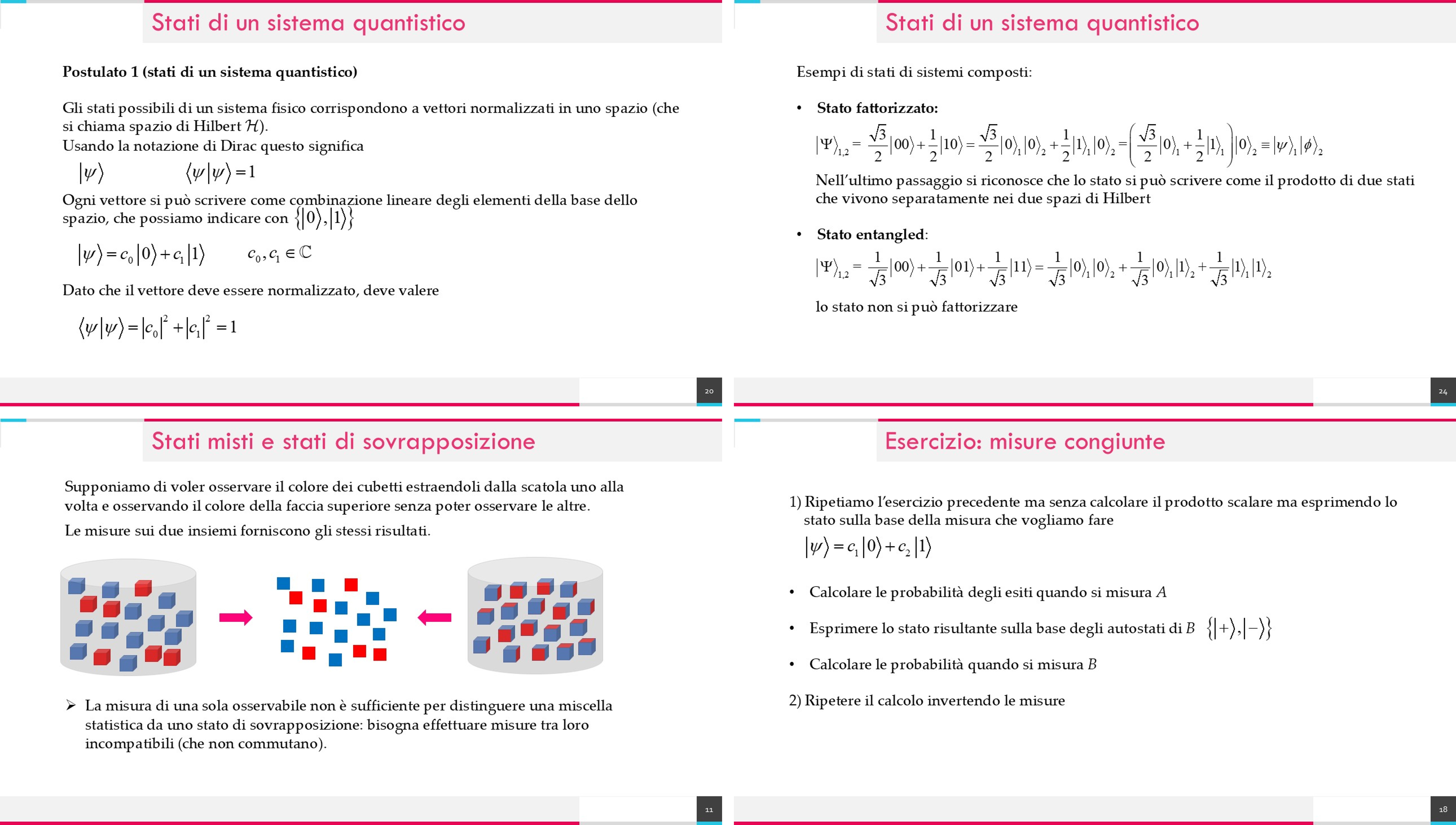}
\vspace{0.2 cm}
\caption{Example of slides extracted from the second and third lectures.}\label{L2_3}
\end{figure}

\begin{figure}[ht]
\includegraphics[width=1\textwidth]{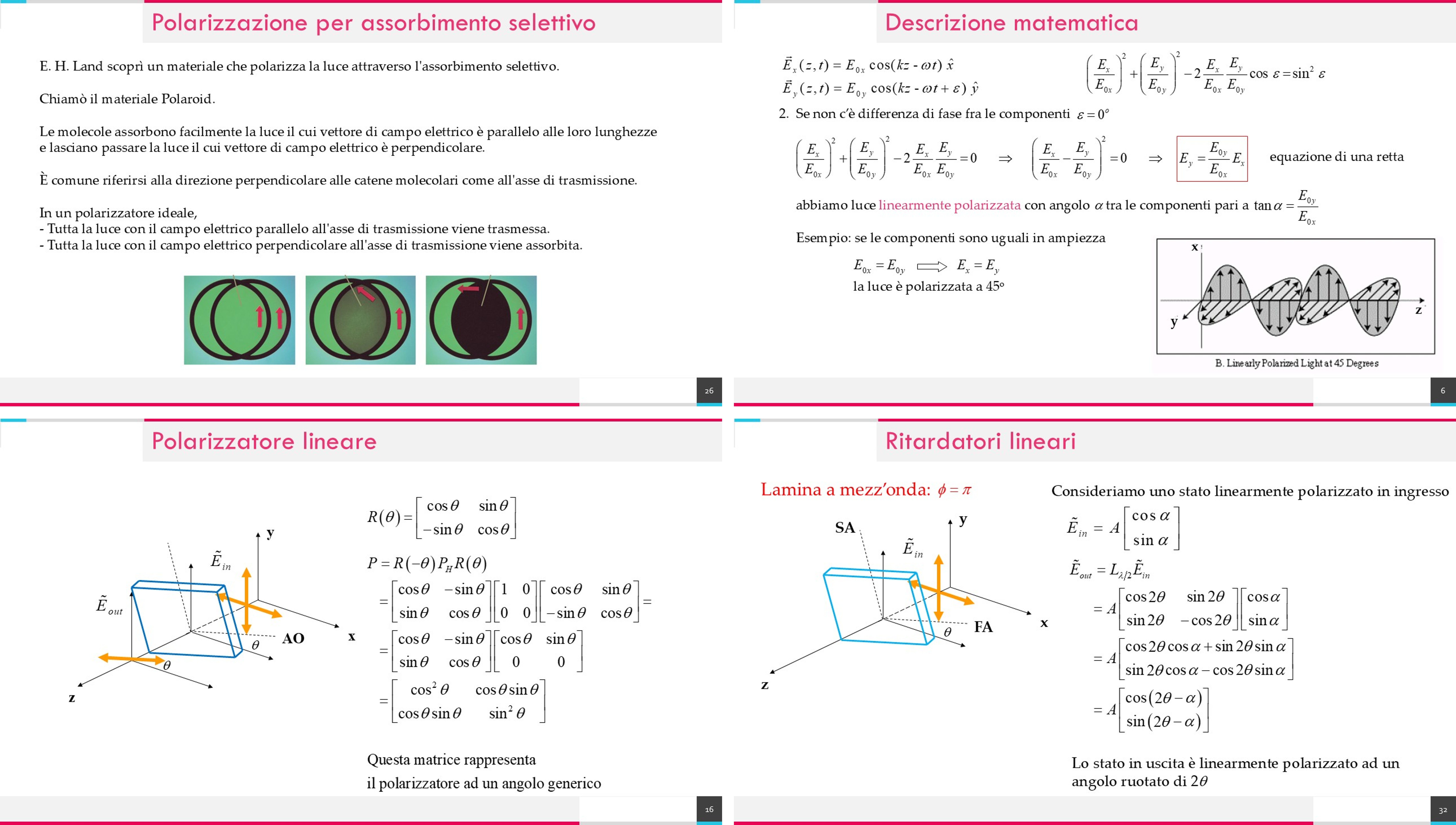}
\vspace{0.2 cm}
\caption{Example of slides extracted from the third and fourth lectures.}\label{L3_4}
\end{figure}

\begin{figure}[ht]
\includegraphics[width=1\textwidth]{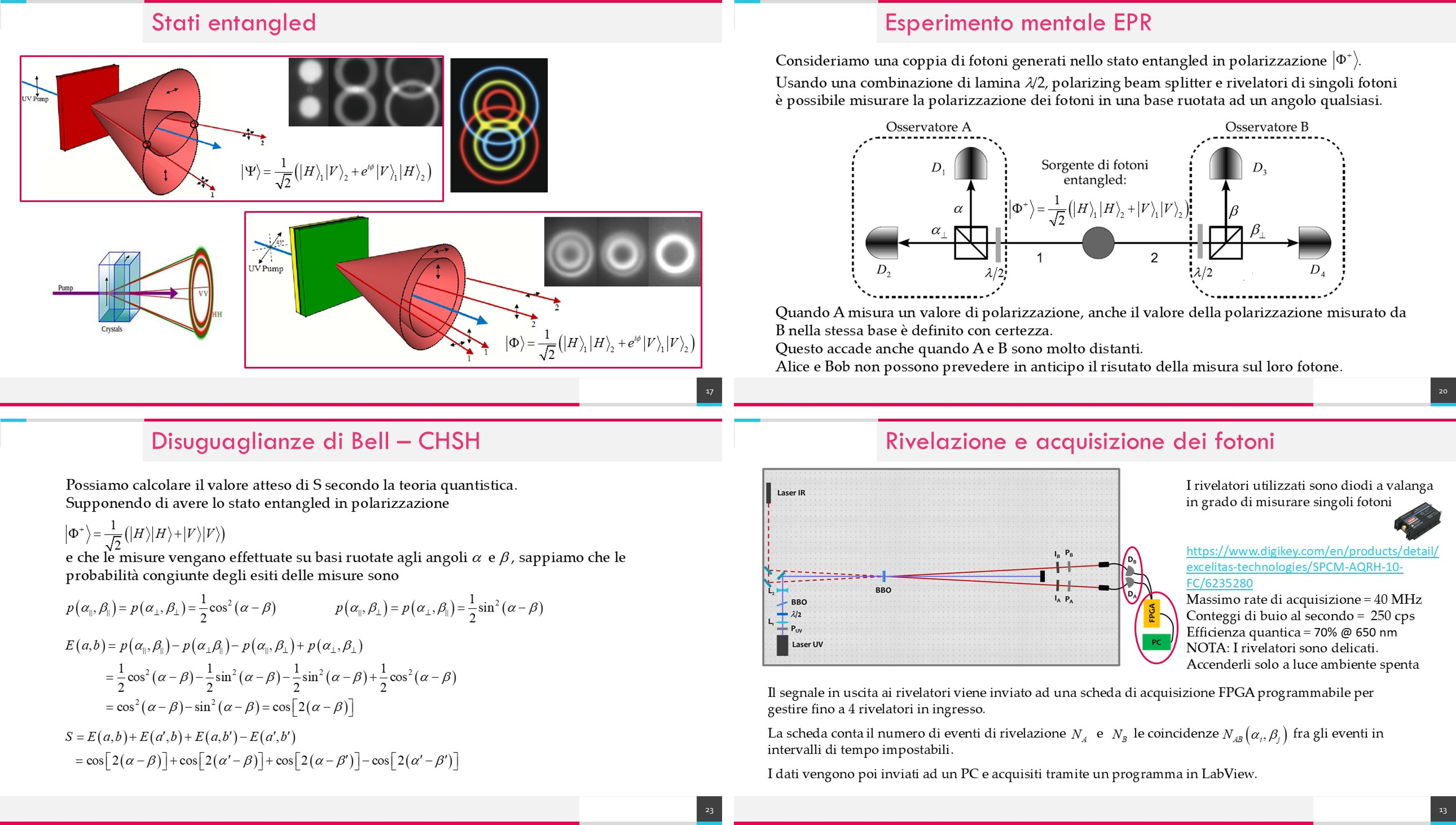}
\vspace{0.2 cm}
\caption{Example of slides extracted from the fifth and sixth lecture.}\label{L5_6}
\end{figure}

This appendix provides a detailed description of the educational activities implemented during the student course in Genoa at the Liceo M.L. King in the spring of 2025. The teacher professional development programme followed the same conceptual progression but devoted additional time to pedagogical discussions and classroom implementation strategies. The student pathway consisted of 12.5 hours of theoretical instruction and 12 hours of laboratory activities, interleaved throughout the course to maintain a close connection between formal concepts and experimental practice. As an example of the lecture materials, some of the original slides (in Italian) are shown in Figs.~\ref{L1_2}--\ref{L5_6}.

\begin{table}[h]
\label{tab_tot}
\caption{Course organization.}
\begin{tabular}{>{\centering\arraybackslash}m{1.5cm}p{6cm}>{\centering\arraybackslash}m{1.5cm}}
\hline\noalign{\smallskip}
\textbf{Module} &
\textbf{Topic} &
\textbf{Hours} \\
\hline
\noalign{\smallskip}
A.1 &
Classical physics and CHSH game.&2.5\\
A.2 &
Mathematical tools and two-state formalism.&4 \\
A.3 &
Polarization as a qubit.&3.5 \\
A.4 &
Entanglement and Bell inequalities.&2.5 \\
A.5 &
Laboratory activities.&12 \\
\noalign{\smallskip}\hline
\end{tabular}
\end{table}

\subsection{Classical description of physical systems}
\subsubsection{Classical states and determinism (1 h)}

The activity began with a discussion on the way in which classical physics describes the physical world. Considering the simple example of a classical moving particle, the idea of the complete description of its motion by assigning position and velocity in the phase space was introduced. This led to explore the concepts of trajectories, state, and initial conditions, stressing that no classical particle can be in two different points at the same time. Moreover, the discussion explored what happens when initial conditions are unknown, leading to the concept of classical probability as a tool to manage ignorance.

\subsubsection{CHSH card game (1 h)}

Then we played a card game designed to explore the concept of correlation and derive the classical limit of the CHSH version of Bell's inequalities \cite{Derenzi2025,Derenzi2026}. Several pairs of card decks were prepared coupling the cards in different ways and then distributed to two student groups A and B, who then made ``measurements” on card properties: color of the back (B), either red or blue and color of the front face (F), either red or black. To each result of a measurement was assigned a value of $ b=\pm 1$ (red/blue) or  $f=\pm 1$ (red/black), allowing the construction of correlation functions for the different pair of properties: $\langle b_A b_B \rangle$, $\langle b_A f_B \rangle$, $\langle f_A b_B \rangle$, $\langle f_A f_B \rangle$. Hence the CHSH function was defined as: $\langle S \rangle = \langle b_A b_B \rangle + \langle b_A f_B \rangle + \langle f_A b_B \rangle - \langle f_A f_B \rangle$. With simple calculations it is easy to demonstrate that $|\langle S \rangle | \leq 2$  for any preparation of the decks \cite{Clauser1969}. Direct measurements make the classical result easy to check. By analyzing the actual CHSH function values for differently prepared pairs of decks, it is also possible to infer information about how the decks were prepared, distinguishing between correlations, anticorrelations, and random behavior.

Beyond introducing Bell's inequalities in an intuitive manner, the game established the concepts of local realism, hidden variables, and statistical correlations that were repeatedly revisited throughout the course.

\subsubsection{Reflection on Bell's inequality assumptions (0.5 h)}

After the game, a reflection took place on the hidden assumptions underlying the result: the existence of predefined properties, the meaning of measurement, the independence of measurements (simulated here by separating the groups), and the notion of correlation. There was also a discussion on the need for an adequate sample size (limited by the number of cards) to ensure statistical significance.

\subsection{Introduction to the two-state formalism}
\subsubsection{Mathematical interlude (1 h)}

The mathematical tools required by the course were introduced through a brief review of vectors, matrices, basis transformations, and complex numbers, directly connected to the physical problems addressed later in the curriculum.

\subsubsection{Quantum states and measurements (3 h)}

The formalism of quantum mechanics was introduced using the Dirac notation and two-state systems. 
The quantum description was then introduced: defining states, the state space, superposition, measurement, and evolution. The key differences from the classical model discussed in the first lecture are highlighted step by step. Everything was developed using the Dirac notation, leveraging the algebraic simplicity of two-level systems as tools to handle operators and transformations \cite{Bondani2022,Pallotta2022,Amabile2026}. The treatment was extended to composite systems, leading to the concept of entanglement, discussed in terms of factorizability of the quantum states. By comparing with the card game, students could see how an entangled state cannot be written as a product of two single-system states. Both the mathematical and physical properties of Bell states were then explored, including analytical calculations of measurements in different bases.

Finally the role of measurement in quantum physics was discussed stressing the existence of incompatible observables that cannot be measured simultaneously with certainty. Particular attention to the representation of the states on different bases was discussed to allow for straightforward implementation of different measurements.

\subsection{Polarization as a physical realization of qubits (3.5 h)}

Since the experiment of violation of Bell’s inequality will be implemented on polarization-entangled states, a lot of time was devoted to introduce and discuss light polarization. The first step was to define what light polarization is, describing light as a polarized electromagnetic wave and making examples on macroscopic phenomena connected to it. Natural and artificial phenomena related to polarization were discussed, along with the operation of polarizers. Then polarization was formalized with a vector description consistent with the Dirac notation introduced earlier. Polarization states were described by 2D vectors while polarizers and wave plates by 2X2 matrices.
As an application of the formalism, the use of wave plates as unitary operators and polarizers as measurement tools was illustrated. The effect of sequences of polarizers is theoretically analyzed and analytically calculated \cite{Pallotta2022}.

Students analyzed sequences of optical elements analytically before reproducing the corresponding experiments in the laboratory, reinforcing the relationship between mathematical predictions and experimental observations.

\subsection{Entanglement and Bell's inequalities (2.5 h)}

The final theoretical module introduced spontaneous parametric down-conversion as the physical mechanism used to generate polarization-entangled photon pairs. 
Students were then introduced to the EPR paradox \cite{Einstein1935} and Bell’s inequalities (in the CHSH form) \cite{Bell1964,Clauser1969}. Linking back to the initial game and Bell’s framework, it was shown how the assumptions of locality and realism must be respected by all classical bipartite systems both correlated and uncorrelated. Then using the formalism of polarization-entangled state and exploiting the rules for measurements the expected violation of CHSH inequality is demonstrated.

\subsection{Laboratory activities}
\subsubsection{Polarization experiments (4 h)}

The experimental sessions started with experiments to get familiar with the properties of polarizers and wave plates. Students were divided into small groups and performed quantitative measurements of light intensity at the output of one polarizer (Malus’s law) and of a series of polarizers and qualitative observations on quarter-wave plates.

Besides illustrating polarization phenomena, these experiments provided practical experience with optical alignment, uncertainty estimation, and comparison between theoretical predictions and experimental measurements.

\subsubsection{Analysis of the Bell apparatus (2 h)}

Students then examined the complete Bell apparatus installed in the school quantum optics laboratory. Every single piece of equipment was described in the mathematical and physical framework introduced in the previous lectures, starting with the production of the entangled state in the nonlinear crystals, passing through state selection and detection and ending with coincidence acquisition.
All the issues connected to apparatus optimization (fine-tuning of the entangled state, correlated-photons collection, background noise, stray light…) were discussed and students learned the procedure to optimize the state by looking at the measured coincidences.
 
\subsubsection{Experimental violation of Bell's inequalities (3 h)}

Students actively performed the Bell experiment using polarization-entangled photon pairs. They optimized the source, selected the measurement bases, ran the LabView program to collect the coincidence counts, deriv correlation functions, and evaluate the CHSH parameter from the experimental data.

Unlike a traditional laboratory demonstration, students participated directly in all stages of the experiment, including apparatus optimization and data acquisition.

\subsubsection{Data analysis and discussion (3 h)}

The final laboratory session was devoted to analyzing the experimental results and comparing the measured CHSH parameter with theoretical predictions. Students discussed experimental uncertainties, possible sources of systematic error, and the implications of the observed violation of Bell's inequalities for the interpretation of quantum mechanics.


\section{Description of the alignement procedure}\label{appB}

The main components used in the setup are:
\begin{itemize}
\item Linearly-polarized continuous wave laser source with wavelength $\lambda = 405$~nm, maximum output power $P=150$~mW.
\item Half-wave plate ($\lambda /2$), to rotate the polarization of the laser before it enters the birefringent crystals. 
\item $\beta$-Barium borate crystals: BBO1 and BBO2. Tilting BBO1 around the vertical axis allows the addition of a phase between the vertical and horizontal polarization components of the pump, which is helpful to create the target state. BBO2 is formed by two adjacent crystals, with mutually orthogonal optical axis, and it’s where the Type I SPDC process happens. Here, the first  of the two crystals converts a $|V\rangle$ photon of the pump into two $|H\rangle |H\rangle$ photons, while the second one converts a  $|H\rangle$ photon of the pump into two $|V\rangle |V\rangle$  photons, always obeying energy and momentum conservation. According to the features of the crystal we used, A and B beams at 810 nm make nominal 3° angles with respect to the pump beam. The presence of two thin (compared with the coherence length of the pump) crystals, combined with the pump polarized at 45° leads to an entangled output state that may be approximated as: $| \Psi \rangle = ( |H \rangle |H \rangle  + e^{i \phi} |V\rangle |V \rangle)/\sqrt{2}$, where the relative phase  depends on the different propagation speed of the polarization components in the two crystals and which can be adjusted tilting BBO1.
\item A and B polarizers, which select the polarization of the photons that will be collected, on the A and B path respectively. They need to be calibrated while setting up the experiment, i.e. defining at which rotation angle they allow vertically polarized photons to pass.
\item Optical collectors A and B, which allow coupling of the signal into an optical fiber directed to the detectors. They are mounted on a micrometric 3D manual mechanical moving stage to fine tune their position.
\item SPAD, single photon avalanche detector, to count the number of photons arriving at the A and B optical collectors. In order to limit spurious counts as much as possible it is important that the collectors are protected from light different from direct SPDC light. Moreover, filters can be inserted to limit the collected photons to the wavelengths of interest. In our case we use SPCM-AQRH-10-FC from Excelitas Technologies, with maximum acquisition rate of 40 MHz, dark counts of 250 cps and quantum efficiency at 650nm of 70\%.
\item FPGA, field programmable gate array, with inputs from the two SPADs. It counts the number of photons arriving at the detectors in a settable amount of time and the coincidence events in a settable time window (a typical value for detecting two photons from the same pair is around 10 ns). In our setup we use DE2-115 Development and education board from Terasic.
\item LabVIEW interface running on the PC, designed to show in real time the number of counts on the detectors and of the coincidences between them. We adapted to our needs the interface provided by Beck in his work \cite{Beck2012}, maintaining its main features. Different acquisitions parameters can be chosen, in order to find the best working conditions in the different steps of the experiment (i.e. short averaging time is useful in the initial optimization phase, where it is important to rapidly understand the effect of the rotation of the different optical components, while longer acquisition time is necessary in the measurement phase, where acquiring enough statistics is crucial to obtain significant results). Moreover, the same interface guides the experimenters into the sequence of the measurements required for testing Bell’s inequalities, specifying each time the rotation angle for the A and B polarizers, collecting the data and calculating the value for the CHSH function S  at the end.
\end{itemize}

\begin{figure}[ht]
\centering
\includegraphics[width=1\textwidth]{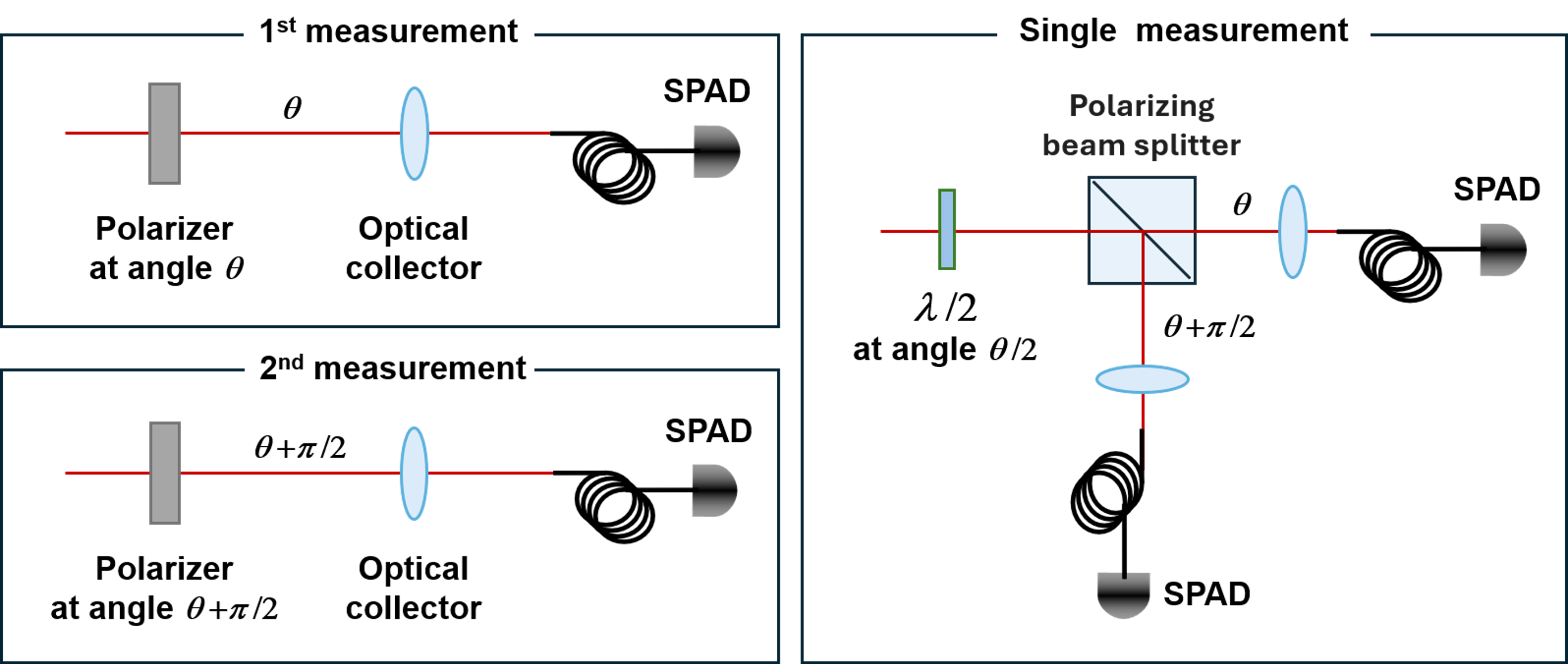}
\vspace{0.2 cm}
\caption{Possible measurement configurations for Alice's and Bob's. Left: measurement of a single polarization component of the state using a single SPAD. Right: simultaneous measurements of the two polarization components of the state using two SPADs.}\label{fig_detector}
\end{figure}

One of the main differences between our experimental set-up and the one presented in \cite{Beck2012} is the number of SPADs employed. Using just two SPADs, in combination with the polarizers, rather than two polarizing beam splitters and four SPADs, significantly reduces the costs of the experimental realization, which is an important aspect in a public school context. However, using the chosen approach the measurement time is multiplied by four, since in this case two sets of measurements are necessary for both Alice and Bob, with the polarizer placed in two orthogonal positions, to reproduce the output of the configuration set by the combination of half-wave plate and polarizing beam splitter. This concept is presented in Fig. \ref{fig_detector}.  

In the following we summarize the main steps for setting up the experiment. 
\begin{itemize}
\item Assembling and rough alignment of the different components, according to the schematic presented in Fig. \ref{fig_experiment}. For this we proceed using geometric tools, positioning two wires on the optical bench along the theoretical directions of the downconverted photons at 810 nm, and placing the components accordingly. 
\item Initial calibration of the optical components (i.e. define the “zero” on the different wave plates and polarizers, corresponding to no linear polarization rotation and transparence to vertically polarized photons respectively). For this scope we use a laser in combination with a polarizing beam splitter (PBS), which separates $|V \rangle$ from $|H \rangle$ components. 
\item Fine alignment of the system. Optical collectors are slightly translated and tilted using their micrometric stages in order to maximize the counts and, even more importantly, the coincidence rate between SPAD A and SPAD B. In this phase the potential presence of spurious light is checked and consequently the optical collectors are properly isolated to reduce it as much as possible.
\item Preparation of Bell’s state. This is the most crucial part for the success of the experiment, and requires several iteration steps to obtain satisfying results. Given the generic BBO2 output state, to obtain the Bell state we need to have an equal amplitude and a zero relative phase for the $| HH \rangle$ and the $| VV \rangle$ components in the entangled state produced by the BBO2. Experimentally, this corresponds to properly optimize the rotation angle of the half-wave plate and the tilting angle of BBO1.This procedure is inspired by the one described in \cite{Beck2012}, to which we address the reader for more technical details. In our conditions (no air conditioning, no active tables,...) it is necessary to repeat these optimization steps before each acquisition, especially if it is performed in a different day respect with the previous one. 
\item Bell’s inequalities testing: in the chosen experimental setting (with only 2 SPADs), we need 16 measurements, each one corresponding to a different configuration of the polarizers in front of the SPADs. The Labview interface guides the experimenters in the data acquisition, specifying each time how the polarizers A and B need to be rotated and recording all the data.
\end{itemize}

\newpage
\section{Qualitative analysis of students' interviews}\label{appC}

This Appendix provides a detailed description of the protocol and the rubric adopted to analyze students' answers. 

\subsection{Interview questions}
\begin{itemize}
  \item Q1: To what extent did the mathematical formalism presented during the course help or hinder your understanding of the physical phenomena?
  \item Q2: Do you think that the EPR paradox and Bell inequalities are culturally important topics that everyone should know about?
  \item Q3: Learning that Bell inequalities are violated, and therefore that locality and realism cannot simultaneously hold, did this create any cognitive or psychological conflict for you?
  \item Q4: Compared with the beginning of the course, do you now feel more confident discussing physics and more capable of learning new scientific concepts?
  \item Q5: Compared with your expectations before the course, how would you evaluate the overall experience?
\end{itemize}

\subsection{Dimensions}
\begin{itemize}
  \item D1: Relationship between mathematical formalism and conceptual understanding;
  \item D2: Perceived cultural and epistemological significance of quantum mechanics;
  \item D3: Students' cognitive responses to the violation of Bell inequalities;
  \item D4: Changes in disciplinary self-efficacy;
  \item D5: Overall educational impact of the learning experience.
\end{itemize}

\newpage
\subsection{Rubric}
The rubric for the analysis is summarized in Table~\ref{rubric}.

\begin{table}[ht]
\tiny
\centering
\resizebox{1\textwidth}{!}{%
\begin{tabular}{p{0.01 cm} p{10.09 cm} p{0.1 cm}}
\hline
&\textbf{D1: Mathematics as a tool for conceptual understanding}&  \\
\end{tabular}%
}
\resizebox{1\textwidth}{!}{%
\begin{tabular}{p{2.7 cm} p{3.8cm} p{3.7cm}}
\hline
\textbf{Level} & \textbf{Description} & \textbf{Observable Indicators}  \\
\hline
L1 – Barrier & Mathematics is perceived as an obstacle to understanding physics. & "Too many calculations"; "I could not understand what the equations meant."\\
L2 – Separation &  Mathematics is considered useful but perceived as separate from physics. & Distinction between manipulating formulas and understanding phenomena.\\
L3 – Support & Mathematics is viewed as an interpretative tool that helps explain physical phenomena. & Explicit links between equations, predictions, and experiments.\\
L4 – Epistemic integration & Mathematics and physical theory are regarded as inseparable components of scientific explanation. & Recognition that mathematical structure constitutes the theory itself.\\
\end{tabular}%
}
\resizebox{1\textwidth}{!}{%
\begin{tabular}{p{0.01 cm} p{10.09 cm} p{0.1 cm}}
\hline
&\textbf{D2: Cultural value of the EPR–Bell problem}&  \\
\end{tabular}%
}
\resizebox{1\textwidth}{!}{%
\begin{tabular}{p{2.7 cm} p{3.8cm} p{3.7cm}}
\hline
\textbf{Level} & \textbf{Description} & \textbf{Observable Indicators}  \\
\hline
L1 – Technical topic & Considered relevant only for specialists. & "Only physicists need to know this."\\
L2 – Disciplinary interest & Important for students interested in physics. & Academic relevance for future studies.\\
L3 – Cultural relevance &  Connected with phylosophy of science. & References to philosophy of science or the nature of reality.\\
L4 – Universal relevance & Considered part of general scientific literacy. & Explicit discussion of implications for society or our worldview.\\
\end{tabular}%
}
\resizebox{1\textwidth}{!}{%
\begin{tabular}{p{0.01 cm} p{10.09  cm} p{0.1 cm}}
\hline
&\textbf{D3: Cognitive impact of Bell's Inequality violations} &  \\
\end{tabular}%
}
\resizebox{1\textwidth}{!}{%
\begin{tabular}{p{2.7 cm} p{3.8cm} p{3.7cm}}
\hline
\textbf{Level} & \textbf{Description} & \textbf{Observable Indicators}  \\
\hline
L1 – No cognitive conflict & Passive acceptance or lack of reflection. & No indication of surprise or conceptual tension.\\
L2 – Initial cognitive discomfort & Recognition that quantum mechanics is counterintuitive.	& Expressions such as "strange," "unexpected," or "hard to imagine."\\
L3 – Conceptual conflict & Explicit awareness of the incompatibility between classical intuition and quantum theory.& 	Comparison between classical and quantum descriptions.\\
L4 – Epistemological restructuring	& Reflection on the meaning of reality, measurement, and scientific knowledge. & Discussion of realism, locality, observation, or the limits of classical concepts.
\end{tabular}%
}
\resizebox{1\textwidth}{!}{%
\begin{tabular}{p{0.01 cm} p{10.09  cm} p{0.1 cm}}
\hline
&\textbf{D4: Disciplinary self-efficacy}&  \\
\end{tabular}%
}
\resizebox{1\textwidth}{!}{%
\begin{tabular}{p{2.7 cm} p{3.8cm} p{3.7cm}}
\hline
\textbf{Level} & \textbf{Description} & \textbf{Observable Indicators}  \\
\hline
L1 – No change & Confidence remains unchanged. & Persistent insecurity.\\
L2 – Limited improvement & Greater confidence only regarding the specific topics studied. & Improved familiarity with Bell's inequalities but not with physics more generally.\\
L3 – Disciplinary growth & Increased confidence in discussing and learning physics.	& Greater curiosity, motivation, and willingness to engage with new concepts.\\
L4 – Identity development & Students begin to perceive themselves as capable learners of advanced physics. & 	Explicit references to autonomy, confidence, and future scientific learning. \\
\end{tabular}%
}
\resizebox{1\textwidth}{!}{%
\begin{tabular}{p{0.01 cm} p{10.09   cm} p{0.1 cm}}
\hline
&\textbf{D5: Overall evaluation of the educational experience}&  \\
\end{tabular}%
}
\resizebox{1\textwidth}{!}{%
\begin{tabular}{p{2.7 cm} p{3.8cm} p{3.7cm}}
\hline
\textbf{Level} & \textbf{Description} & \textbf{Observable Indicators}  \\
\hline
L1 – Negative mismatch & Expectations not fulfilled; course perceived negatively. & Disappointment or frustration.\\
L2 – Mixed evaluation & Positive and negative aspects coexist.	 & Generally positive assessment despite major reservations.\\
L3 – Positive evaluation & Course regarded as valuable and intellectually stimulating. & Satisfaction with learning outcomes.\\
L4 – Transformative experience & Students report a meaningful change in their understanding of physics or science. & Explicit statements describing a new way of thinking about physics.\\
\hline
\end{tabular}%
}
\caption{Rubric for the analysis of students' interviews.}
\label{rubric}
\end{table}

\end{appendices}


\end{document}